\documentclass[%
 reprint,
 amsmath,amssymb,
 aps,
 prl,
]{revtex4-2}

\usepackage{graphicx}
\usepackage{dcolumn}
\usepackage{bm}
\usepackage{siunitx}
\usepackage{braket}
\usepackage{xcolor}

\usepackage{hyperref}
\hypersetup{
    colorlinks,
    linkcolor={black},
    citecolor={blue!80!black},
    urlcolor={blue!80!black}
}

\begin{document}

\title{Collective excitation and decay of waveguide-coupled atoms:\\from timed Dicke states to inverted ensembles}

\author{Christian Liedl}
\author{Sebastian Pucher}%
\author{Felix Tebbenjohanns}%
\author{Philipp Schneeweiss}%
\author{Arno Rauschenbeutel}%
 \email{arno.rauschenbeutel@hu-berlin.de}
\affiliation{
Department of Physics, Humboldt-Universität zu Berlin, 10099 Berlin, Germany
}

\date{\today}

\begin{abstract}
The collective absorption and emission of light by an ensemble of atoms is at the heart of many fundamental quantum optical effects and the basis for numerous applications. However, beyond weak excitation, both experiment and theory become increasingly challenging. Here, we explore the regimes from weak excitation to inversion with ensembles of up to one thousand atoms that are trapped and optically interfaced using the evanescent field surrounding an optical nanofiber. We realize strong inversion, with about 80~\% of the atoms being excited, and study their subsequent radiative decay into the guided modes. The data is very well described by a simple model that assumes a cascaded interaction of the guided light with the atoms. Our results contribute to the fundamental understanding of the collective interaction of light and matter and are relevant for applications ranging from quantum memories to sources of nonclassical light to optical frequency standards.
\end{abstract}

\maketitle

The collective emission of radiation by an ensemble of atoms is a central problem in quantum optics that has recently seen renewed interest in the context of optical quantum technologies. There, the super- and subradiant states that emerge when a propagating light field couples to a spatially extended ensemble of emitters can be used as a resource to implement novel protocols, e.g., in the context of quantum information, quantum communication, and frequency standards~\cite{chang2018colloquium, duan2001long, meiser2009prospects, facchinetti2016storing, asenjo2017exponential}. The underlying, so-called timed Dicke regime has been extensively studied, both theoretically and experimentally, e.g., with ensembles of laser-cooled atoms in the optical domain~\cite{scully2009super, guerin2016subradiance, araujo2016superradiance, roof2016observation, das2020subradiance, doespiritosanto2020collective, rui2020subradiant, he2020geometric, stiesdal2020observation}. However, the regime where the ensemble is highly excited or even fully inverted has only recently become accessible~\cite{ferioli2021storage, cipris2021subradiance, ferioli2021laser-driven, glicenstein2022superradiance}. There, the theoretical description becomes increasingly complex due to the exponential scaling of the system's Hilbert space with the number of emitters~\cite{masson2020many, masson2021universality, sierra2021dicke, robicheaux2021theoretical, rubiesbigorda2021superradiance, lemberger2021radiation, plankensteiner2022quantumcumulants}. Recently, nanofiber-based atom-light interfaces have opened a new experimental avenue for studying collective radiative dynamics with waveguide-coupled atoms~\cite{nayak2018nanofiber, sheremet2021waveguide, goban2015superradiance, solano2017super, corzo2019waveguide, okaba2019superradiance, pennetta2021observation, pennetta2022collective}. There, all emitters couple efficiently to the guided optical mode, and propagation-direction-dependent coupling can be implemented, providing access to the field of chiral quantum optics~\cite{lodahl2017chiral}. Moreover, the waveguide mediates an effectively infinite-range interaction between macroscopically separated atoms~\cite{lekien2005nanofiber, zoubi2010metastability, shahmoon2013nonradiative, gonzalez2015chiral, solano2017super}.  

Here we explore the collective radiation of an ensemble of up to a thousand waveguide-coupled atoms from weak excitation to almost full inversion. The atoms are trapped close to an optical nanofiber and excited by a resonant, fiber-guided probe pulse that is much shorter than the excited state lifetime. We observe Rabi oscillations of the ensemble by counting the number of photons absorbed from the excitation pulse and achieve almost full inversion. Following the excitation, we measure the fluorescence emitted into the nanofiber-guided modes, infer a collective coupling efficiency, and study its dependence on the pulse area and on the number of atoms. We compare our data to the predictions of a simple model that assumes unidirectional coupling and, consequently, a cascaded interaction of the nanofiber-guided light with the atoms and find excellent agreement.

\begin{figure}
  \centering
  \includegraphics[width=0.98\columnwidth]{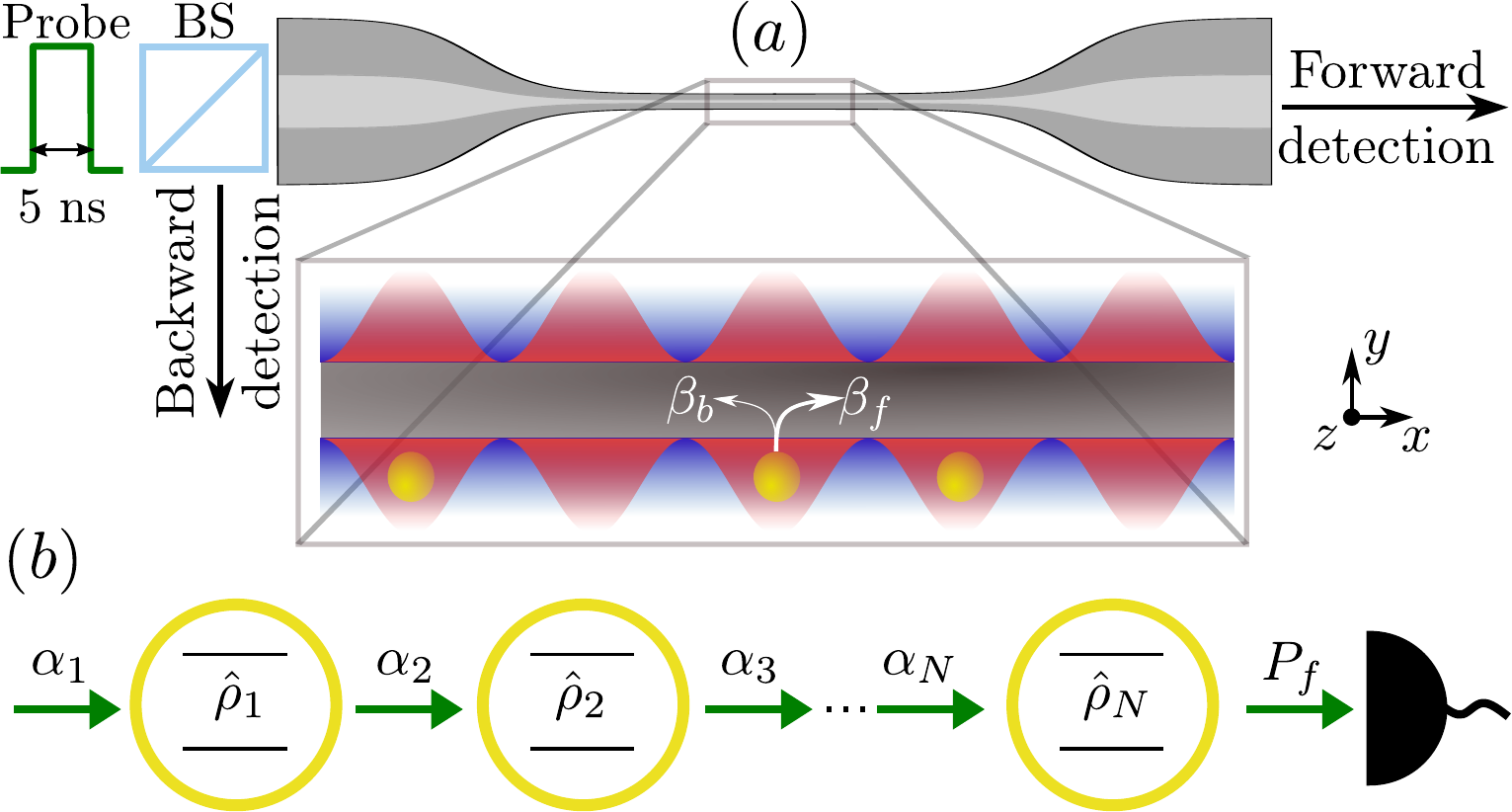}
\caption{(a) Schematic of the experimental setup. Cesium atoms (yellow discs) in a nanofiber-based dipole trap are evanescently interfaced with a guided probe pulse. The atom-waveguide coupling strengths $\beta_f$ and $\beta_b$ are propagation-direction-dependent, indicated by the white arrows. 
The transmitted and reflected pulses are sent to two detection setups. (b) We model the system dynamics using a cascaded interaction model, where the $k$th atom is described by the density operator $\hat{\rho}_k$ and is driven by a coherent field with amplitude $\alpha_k$.}
\label{fig:setup}
\end{figure}

Our experimental setup is schematically shown in Fig.~\ref{fig:setup}(a). The nanofiber is realized as the waist of a tapered optical fiber (TOF) and has a nominal diameter of $\SI{500}{nm}$. A running-wave blue-detuned nanofiber-guided field (wavelength $\SI{760}{nm}$, power $\SI{20.5}{mW}$) and a standing-wave red-detuned field (wavelength $\SI{1064}{nm}$, total power $\SI{2.4}{mW}$) form two diametral arrays of optical trapping sites, located about $\SI{230}{nm}$ from the nanofiber surface. We prepare cesium atoms on only one side of the fiber with at most one atom per trapping minimum~\cite{schlosser2002collisional, vetsch2010optical, meng2018near, supplemental}. We then infer the number of trapped atoms, $N$, using transmission spectroscopy~\cite{vetsch2010optical}.
Throughout the experimental sequence described below, we continuously perform degenerate Raman cooling (DRC) on the  $\ket{6S_{1/2}, F=4}\to\ket{6P_{1/2}, F=4}$  D1-transition using a free-space laser field with a scattering rate that is far smaller than the collective decay rate of the atoms, thus not altering the collective dynamics. Moreover, DRC continuously pumps the atoms to the outermost Zeeman ground state, $\ket{6S_{1/2}, F=4, m_F=-4}$.
 
In order to study the collective dynamics, we launch probe light pulses into the fiber, which are resonant with the $\ket{6S_{1/2}, F=4}\to\ket{6P_{3/2}, F=5}$ D2 transition. 
At the position of the atoms, the probe light is almost perfectly $\sigma^-$-polarized with respect to the quantization axis along $+z$~\cite{lekien2004field}. It thus predominantly drives the cycling transition between ground state $\ket{g}=\ket{6S_{1/2}, F=4, m_F=-4}$ and excited state $\ket{e}=\ket{6P_{3/2}, F=5, m_F=-5}$, effectively realizing a two-level atom. We derive the probe pulses from a continuous wave laser using external amplitude modulation with two cascaded Mach-Zehnder switches. For all measurements, the pulse length is set to $T_\text{pulse}=\SI{5}{ns}$, i.e. much shorter than the $6P_{3/2}$ state's natural lifetime of about 30.5~ns~\cite{steck}. Per sequence, we launch 400 probe pulses into the nanofiber (repetition rate $\SI{5}{\kilo\hertz}$) and record time traces of the output power in both forward and backward direction using single-photon counting modules, see Supplemental Material~\cite{supplemental}. The number of trapped atoms stays essentially constant during this probing (we measure that at most $\SI{15}{\percent}$ are lost). To obtain sufficient counting statistics, we average the recorded traces over several hundred sequences.

\begin{figure}
  \centering
  \includegraphics[width=0.98\columnwidth]{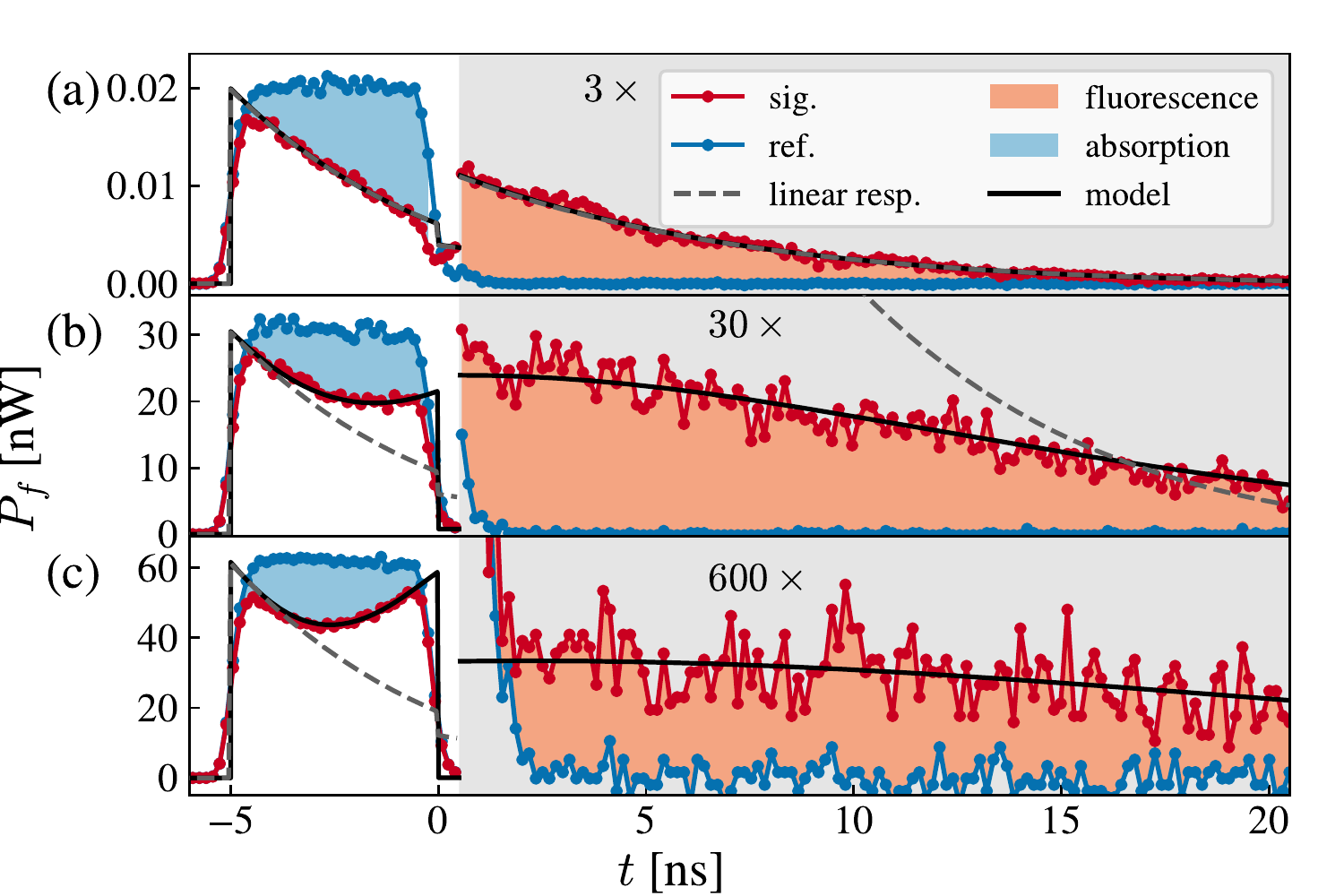}
\caption{Dynamics of the power of a fiber-guided, resonant $\SI{5}{\nano\second}$ probe pulse transmitted through about $300$ atoms for different input powers (red dots). The reference measurement without atoms is shown as blue dots. (a) In the weak excitation regime, the dynamics is well captured by linear response theory (gray dashed line). (b), (c) For larger powers, the dynamics deviates from linear response, but is captured by a cascaded interaction model (black solid line). We extract the number of photons absorbed and emitted into the forward propagating mode from the area of the blue and light-red shaded areas, respectively. Note the rescaled y-axis for $t>0$~ns.}
\label{fig:time_traces}
\end{figure}

Figure~\ref{fig:time_traces} shows time traces (red dots) of the measured output power in the forward direction, $P_f$, for different input pulse powers and about 300 trapped atoms. For comparison, we also show $P_f$ without atoms (blue dots), as well as the prediction of linear response theory, which models the atoms as classical Lorentz oscillators (gray dashed line).  
For the input probe power of 20~pW in Fig.~\ref{fig:time_traces}(a), we observe that $P_f$ decreases during the excitation pulse as predicted by linear response. Subsequently, the atoms emit fluorescence into the waveguide with a collectively enhanced decay constant of 6.1(1)~ns~\cite{pennetta2022collective}.

When we increase the input power by about three orders of magnitude to 30~nW and 60~nW in Figs.~\ref{fig:time_traces}(b) and (c), respectively, the dynamics deviates from linear response. Generally, the strength of the optical drive can be quantified by the Rabi frequency seen by the first atom 
\begin{equation}
\Omega_1=\sqrt{\frac{4\beta_f\Gamma P_1}{\hbar\omega}},
\label{eq:Rabi_freq}
\end{equation}
where $P_1$ is the optical input probe power, $\hbar\omega$ the photon energy, and $\beta_f=\Gamma_f/\Gamma\approx0.01$ the coupling strength to the forward-propagating mode~\cite{supplemental}. Here, $\Gamma_f$ and $\Gamma$ are the atomic emission rates into the forward-propagating mode and free space, respectively. 
The pulse area seen by the first atom is then $A_1=\Omega_1 T_\text{pulse} \approx 0.02 \pi$ in Fig.~\ref{fig:time_traces}(a), i.e. much smaller than $\pi$. Therefore, the atoms reside mostly in their ground state, and the dynamics can indeed be described by linear response theory. 
The pulse areas in Figs.~\ref{fig:time_traces}(b) and (c) are $0.7\pi$ and $\pi$, respectively, such that the nonlinearity of the atoms becomes apparent. In particular, we observe an increase of $P_f$ after an initial decrease in Fig.~\ref{fig:time_traces}(c), which is expected when the ensemble is inverted. The fluorescent decay dynamics for $t>0$~ns also deviates from linear response, with a fluorescence power that is orders of magnitude lower than predicted, see Figs.~\ref{fig:time_traces}(b) and (c).
 
Let us now try to understand the observed nonlinear dynamics quantitatively. Due to the chiral coupling of the atoms to the nanofiber mode, the $\sigma^-$-polarized atomic fluorescence is predominantly emitted into the forward direction~\cite{mitsch2014quantum}. As a consequence, we can describe our system using a cascaded interaction model, see Fig.~\ref{fig:setup}(b). We neglect free-space coupling between the atoms because their average nearest-neighbor distance is larger than half the free-space wavelength of the emitted radiation. We describe the $k$th atom by the density operator $\hat{\rho}_k$ and assume that it is driven by a coherent field with amplitude $\alpha_k\in {\rm I\!R}$. We determine $\alpha_{k+1}$ by interfering $\alpha_k$ with the coherent part of the light field that is emitted into the waveguide by the $k$th atom, yielding the input-output equation~\cite{gardiner1985input}
\begin{equation}
\alpha_{k+1}=\alpha_{k}-i\sqrt{\beta_f \Gamma}\rho_k^{ge}.
\label{eq:input_output}
\end{equation}
Consecutively solving the Lindblad master equation for each atom and Eq.~(\ref{eq:input_output}) for all fields, we obtain the time-dependent quantities $\alpha_k$, $\rho_k^{ge}$, and $\rho_k^{ee}$ and compute the transmitted power, $P_f$, via
\begin{equation}
P_f=P_1+ \hbar\omega\sum_{k=1}^N\big[{\beta_f \Gamma\rho_k^{ee}}+2\sqrt{\beta_f \Gamma}\Im\{\alpha_k^*\rho_k^{ge}\}\big],
\label{eq:power}
\end{equation}
where $\Im\{\cdot\}$ denotes the imaginary part and $^*$ complex conjugation. In general, $P_f$ differs from $\hbar\omega|\alpha_{N+1}|^2$ because the latter only describes the coherent part of the transmitted radiation while the former also accounts for the incoherent part given by the terms proportional to $\rho_k^{ee}$ in Eq.~(\ref{eq:power}). We note that our assumption that the atoms are exclusively driven by the coherent part of the impinging light fields is not justified, e.g., for large ensembles in a fully inverted state. To account for temperature-induced fluctuations of the atomic positions and the corresponding coupling strengths, we average over a Gaussian distribution of $\beta_f$ values. We then fit the mean and standard deviation to our experimental data, yielding 0.0108 and 0.0065, respectively, see Supplemental Material~\cite{supplemental}. This fitted mean value of $\beta_f$ agrees reasonably with an independent saturation measurement~\cite{vetsch2010optical} that yields $\beta_f=0.009(1)$. The predicted power according to Eq.~(\ref{eq:power}) is shown as black solid lines in Fig.~\ref{fig:time_traces}(a)--(c). In the weak excitation regime, it reproduces the predictions obtained by linear response theory, see Fig.~\ref{fig:time_traces}(a). Remarkably, the model also describes the dynamics in the nonlinear regime well. 

\begin{figure}
  \centering
  \includegraphics[width=0.98\columnwidth]{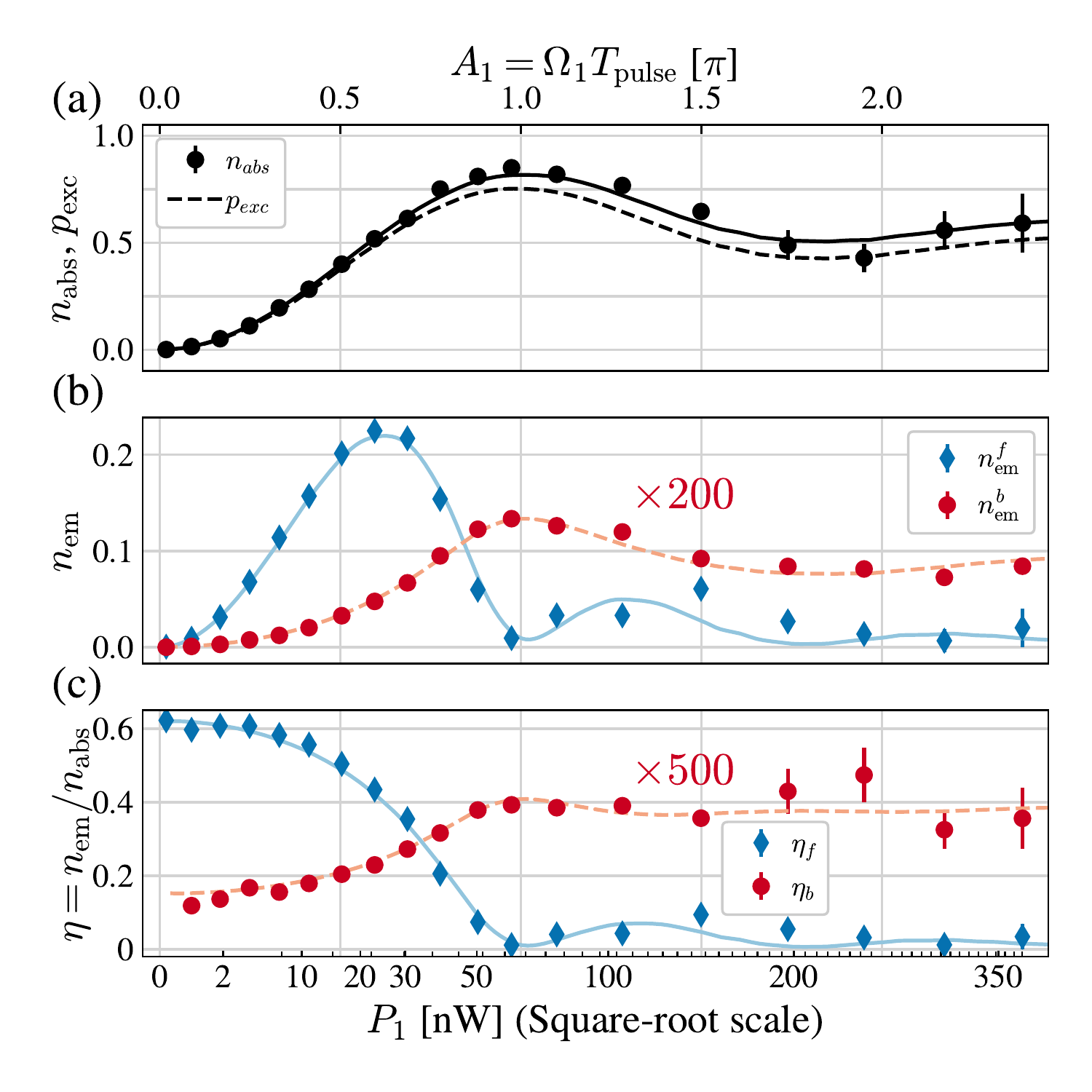}
\caption{(a) Measured average number of absorbed photons per atom, $n_\text{abs}$, as a function of the input power, $P_1$, and the corresponding pulse area seen by the first atom, $A_1$. The black solid and dashed lines show the model prediction for $n_\text{abs}$ and the excited state probability, $p_\text{exc}$, respectively. We observe damped Rabi oscillations, where up to about 80~$\%$ of the atoms are being excited. (b) Number of photons per atom emitted into the forward (backward) propagating mode, $n_\text{em}^f$ ($n_\text{em}^b$), shown as blue diamonds (red dots). The corresponding model prediction is indicated by the blue solid (red dashed) line. (c) Fraction of stored energy emitted into the forward (backward) propagating mode, $\eta_f$ ($\eta_b$). For weak driving, $\eta_f$ is collectively enhanced. As the power is increased,  $\eta_f$ decreases and reaches its minimal value for a pulse area of $\pi$, while $\eta_b$ follows the opposite trend.}
	\label{fig:beta_vs_Omega}
\end{figure}

To further shed light on the system dynamics, we determine the absorbed pulse energy from the blue shaded areas in Fig.~\ref{fig:time_traces}. From this, we infer the average number of absorbed photons per atom, $n_\text{abs}$, which we show in Fig.~\ref{fig:beta_vs_Omega}(a) as a function of the input power $P_1$ (and pulse area $A_1$). The black dots indicate the measured values, and the solid black line our model prediction, which agrees quantitatively with the data. The corresponding model prediction of the average excitation probability at $t=0$, $p_\text{exc}$, is shown as the dashed black line in Fig.~\ref{fig:beta_vs_Omega}(a). We observe damped Rabi oscillations, where $n_\text{abs}$ and $p_\text{exc}$ reach maximum values of 0.85(1) and 0.76(1), respectively. We thus conclude that we achieve significant inversion of the ensemble. During the short probe pulse duration, the probability of an excited atom to decay is small and, thus, $p_\text{exc}$ is slightly smaller than  $n_\text{abs}$. In the following, we use $n_\text{abs}$ as an estimate of $p_\text{exc}$.

We now study the number of photons, $n_\text{em}^f$, that are emitted per atom into the forward direction after switching off the pulse, inferred from the light-red shaded areas in Fig.~\ref{fig:time_traces}. Figure \ref{fig:beta_vs_Omega}(b) shows the measured values of $n_\text{em}^f$ (blue diamonds) and the corresponding model prediction (solid blue line). The maximum value of $n_\text{em}^f$ is as high as 0.225(2), meaning that, on average, about 70 photons are emitted into the forward direction by 300 atoms. Notably, the maximum of $n_\text{em}^f$ does not occur for the same pulse area as the maximum of $p_\text{exc}$, because $n_\text{em}^f$ depends on both the collective enhancement of radiation into the forward propagating mode and the total number of photons stored in the ensemble. While the former stems from constructive interference of the fields radiated by the induced atomic dipoles and is maximized for vanishing input power, the latter peaks for a $\pi$-pulse.

In order to quantify the collective enhancement in forward emission, we compute the fraction of stored energy emitted into the forward direction, $\eta_f=n_\text{em}^f/n_\text{abs}$, shown as blue diamonds in Fig.~\ref{fig:beta_vs_Omega}(c). The solid blue line shows the corresponding model prediction. For vanishing input power, where the atomic emission is fully coherent, the system behaves as a phased array of classical dipoles, featuring enhanced forward scattering, as predicted by linear response. 
There, $\eta_f$ reaches a maximum value of $0.62(1)$, which corresponds to a 60-fold enhancement compared to independent emission. As we increase the input power, incoherent spontaneous emission starts to dominate the decay dynamics. Concomitantly, $\eta_f$ drops, reaching a minimal value of $0.011(2)$ for a pulse area of~$\pi$.

In addition to $P_f$, we also measure the output power in the backward direction, $P_b$. We plot the average number of 
photons per atom, $n^b_\text{em}$, as well as the fraction of stored energy that is emitted into the backward direction,  $\eta_b$, as red dots in Fig.~\ref{fig:beta_vs_Omega}(b) and (c), respectively. We observe that, in contrast to the forward direction, $n_\text{em}^b$ follows a similar trend as $p_\text{exc}$ and peaks at a pulse area of about~$\pi$. We note that, while the model described above assumes unidirectional coupling of the atomic emission into the forward direction, in the actual experimental setting we still expect non-vanishing coupling to the backward-propagating mode, with $\beta_b = 0.087\beta_f$~\cite{mitsch2014quantum}. In order to extract $n_\text{em}^b$ from the cascaded interaction model, we incoherently sum up the time-dependent atomic emission rates to calculate the power emitted into the backward direction
\begin{equation}
P_b = \hbar\omega\sum_{k=1}^N \beta_b\Gamma \rho^{ee}_k,
\label{eq:power_b}
\end{equation}
which we integrate over time to obtain $n_\text{em}^b$.
The assumption of incoherent summation is motivated by the fact that the spatial period of the trapping potential does not fulfill the condition for Bragg-reflection of the probe light, and the atoms are randomly distributed over the trapping sites with a non-unity filling factor. Indeed, the resulting model prediction agrees quantitatively with the experimental data, see dashed red lines in \mbox{Fig.~\ref{fig:beta_vs_Omega}(b) and (c)}. 
\begin{figure}
  \centering
	\includegraphics[width=0.98\columnwidth]{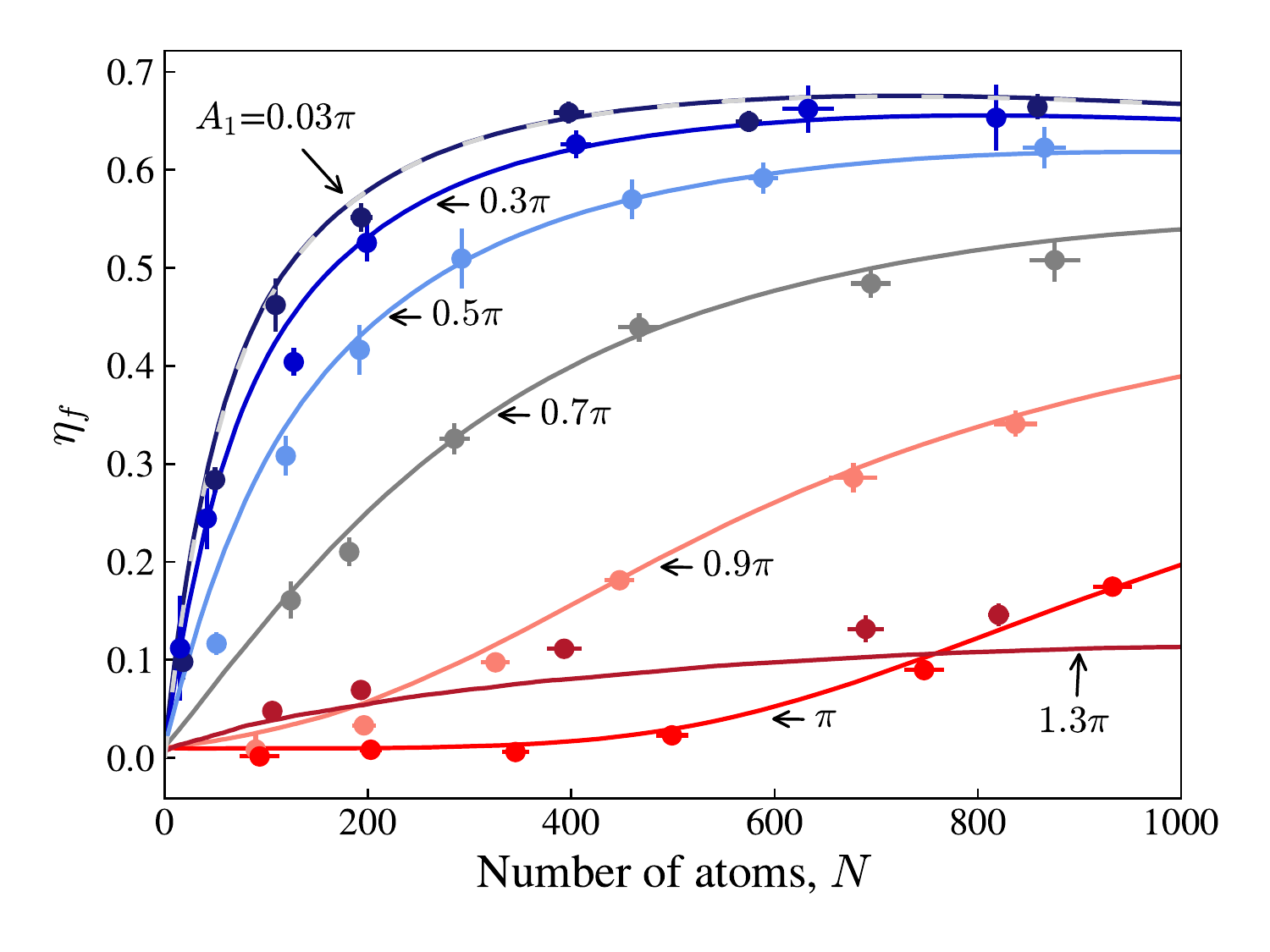}
	\caption{Fraction of stored energy emitted into the forward direction, $\eta_f$, as a function of the atom number, $N$, for different pulse areas, $A_1$. The data is well described by the cascaded interaction model (solid lines). As a comparison, we also plot the result from linear response theory (gray dashed line), which coincides with the prediction of the cascaded interaction model for the smallest pulse area, $A_1=0.03\pi$.}
	\label{fig:fig4}
\end{figure}

In Fig.~\ref{fig:fig4}, we study $\eta_f$ as a function of the atom number, $N$, for different pulse areas $A_1$. The colored dots show the experimental data, while the solid lines are the predictions of the cascaded interaction model, which agrees with the data. For the smallest values of $A_1$, $\eta_f$ first increases linearly with $N$ and then reaches a plateau, where up to 66(2)~\% of the stored energy is emitted into the forward propagating mode. This behavior is well captured by linear response theory, see dashed gray line. Beyond the regime of linear response, both the initial slope of $\eta_f(N)$ and the value of the plateau decrease with $A_1$. For $A_1=0.9\pi$, we observe a qualitative change in the behavior of $\eta_f$. Here,  $\eta_f$ initially increases superlinearly with $N$, which is indicative of the build-up of a collective dipole moment along the array of atoms. For $A_1=\pi$, $\eta_f$ stays at a low constant value until $N\approx 400$, from when on it increases superlinearly. We numerically checked that the initially constant $\eta_f(N)$ is due to the finite temperature of the atoms, see Supplemental Material~\cite{supplemental}. For an even larger pulse area of $A_1=1.3\pi$, $\eta_f$ scales qualitatively similarly with $N$ as for the case of $A_1=0.7\pi$. The smaller slope and plateau value are predominantly due to the larger thermal fluctuations of the pulse areas seen by the different atoms. 

Our work shows that the excellent controllability of waveguide-coupled atomic arrays facilitates the study of collective radiative effects from weak excitation to full inversion. The unidirectional coupling granted by the nanophotonic waveguide allowed us to realize a cascaded system, where the dynamics of each atom depends only on the dynamics of the upstream atoms~\cite{gardiner1993driving, carmichael1993quantum}. This unidirectional coupling, in conjunction with the assumption that the atoms are solely driven by the coherent part of the guided light field, reduces the computational complexity of the model from exponential to linear in the number of atoms. In spite of these simplifications, our model predictions quantitatively agree with the experimental data over the large range of parameters covered here. 

Beyond this parameter regime, interesting avenues for future research open up. For example, for an even larger number of fully inverted atoms, one expects the dynamics of a given atom to be predominantly driven by photons that are spontaneously emitted by the other atoms. In this case, an initial build-up of the collective emission intensity has been predicted~\cite{lekien2008cooperative}, which is analogous to the superradiant burst described in Dicke's seminal work~\cite{dicke1954coherence, gross1982superradiance}. This dynamics occurs even for macroscopically separated emitters and should be experimentally accessible with our system. Additionally, by tuning the directionality of atom-waveguide coupling, we should be able to study dynamic mode competition in the presence of a directional bias. In this case, even a small asymmetry in the coupling strength may lead to a collective build-up of strongly directional emission~\cite{robicheaux2021theoretical}. Finally, our results constitute an important step toward full coherent optical control of atoms that interact via a nanophotonic waveguide, enabling, e.g., the generation of multi-photon quantum states of light~\cite{gonzalez2015deterministic, mahmoodian2020dynamics}.\\

\begin{acknowledgments}
We thank R. Pennetta and J. Volz for stimulating discussions and helpful comments. We acknowledge funding by the Alexander von Humboldt Foundation in the framework of the Alexander von Humboldt Professorship endowed by the Federal Ministry of Education and Research. Moreover, financial support from the European Union's Horizon 2020 research and innovation program under grant agreement No. 800942 (ErBeStA) is gratefully acknowledged.
\end{acknowledgments}

\bibliography{bibliography}
\newpage
\setcounter{equation}{0} 
\setcounter{figure}{0} 
\renewcommand{\theequation}{S\arabic{equation}}
\renewcommand{\thefigure}{S\arabic{figure}}

\onecolumngrid
\section{Supplemental Material}
\subsection{Loading of the nanofiber-based atom trap}
We transfer cesium atoms from a magneto-optical trap into the nanofiber-based trapping potential, consisting of two diametral arrays of trapping sites, via an optical molasses stage \cite{vetsch2010optical}. Due to the collisional blockade effect, there is at most one atom per trapping site~\cite{schlosser2002collisional}. Then, we apply a homogenous magnetic offset field of about $\SI{0.5}{G}$ along $+z$, see Fig.~1(a) in the main text, and further cool the atoms on one side of the nanofiber by degenerate Raman cooling using nanofiber-guided light that is near-resonant with the $\ket{6S_{1/2}, F = 4}\to \ket{6P_{3/2}, F = 5}$ D2-transition~\cite{meng2018near}. In this step, the atoms on the other side of the nanofiber are heated and expelled from the trap, such that a one-dimensional array of atoms on only one side of the nanofiber remains.

\subsection{Detection setup}
Before launching the probe laser field into the tapered optical fiber, we send it through a non-polarizing beamsplitter that reflects 90~$\%$ of the light. This allows us to also detect the light that is backscattered by the atoms, see Fig.~1(a) in the main text. Both in forward and backward direction, the light is spectrally filtered prior to detection to suppress the trapping light. The spectral filtering is designed such that only light with a wavelength around $852$~nm is transmitted. It consists of a dichroic mirror, a volume Bragg grating, and a bandpass filter. 
In the forward direction, the transmitted probe field is split by a non-polarizing $50:50$ beam splitter, and the outputs of the two ports are detected by two single-photon counting modules (SPCM). We use neutral density filters to attenuate the signal incident on one SPCM such that it is not saturated by the strong probe pulses used in our experiment. The signal incident on the other SPCM is not attenuated and measures the transmission spectrum of a weak second probe field in order to infer the number of trapped atoms \cite{vetsch2010optical}. This second SPCM is blocked by a mechanical shutter during the probing sequence to protect it from damage. In the backward direction, the signal field is split by a polarizing beamsplitter and detected by two SPCMs. For the data shown in the main text, the counts from both SPCMs in backward direction are summed up. 

\subsection{Single-atom master equation}
In the cascaded interaction model, the light field is propagated through the ensemble by repeatedly applying the input-output equation, see Eq.~(2) of the main manuscript. For this, we consecutively compute the atomic dynamics for each atom. We describe the $k$th two-level atom with ground state $\ket{g}_k$ and excited state $\ket{e}_k$ by the density operator $\hat{\rho}_k$ and solve the single-atom Lindblad master equation after applying the rotating wave approximation 

\begin{equation}
\frac{\text{d}}{\text{d}t}\hat{\rho}_k = -\frac{i}{\hbar}[\hat{\mathcal{H}_k}, \hat{\rho}_k]+\Gamma\hat{\mathcal{L}}_k.
\end{equation}

Here, $\hat{\mathcal{H}}_k$ is a semi-classical Hamiltonian assuming that the $k$th atom is resonantly driven by a coherent field with a time-dependent Rabi frequency $\Omega_k$, and $\hat{\mathcal{L}}_k$ is the Lindblad superoperator describing spontaneous emission with an excited state population decay rate $\Gamma$

\begin{align}
\hat{\mathcal{H}}_k &= \frac{\hbar\Omega_k}{2}(\hat{\sigma}_k^\dagger+\hat{\sigma}_k)\\
\Omega_k &= \sqrt{4\beta_\text{f}\Gamma}\alpha_k\\
\hat{\mathcal{L}}_k &= \hat{\sigma}_k\hat{\rho}\hat{\sigma}^\dagger_k-\frac{1}{2}\big(\hat{\sigma}^\dagger_k\hat{\sigma}_k\hat{\rho}+\hat{\rho}\hat{\sigma}^\dagger_k\hat{\sigma}_k\big).
\end{align}

Here, $\hat{\sigma}_k=|g\rangle_k\langle e|_k$ is the atomic lowering operator of the $k$th atom, $\beta_f$ the coupling strength of the atoms to the forward-propagating mode, and $\hbar\omega$ the photon energy. For the first atom, the field amplitude $\alpha_1=\sqrt{P_1/\hbar\omega}$ is determined by the time-dependent input probe power, $P_1$, while $\alpha_k$ is obtained from the input-output equation in Eq.~(2) of the main text for atoms $2\ldots N$. 

\subsection{Fluctuations of the coupling strength due to temperature}
\begin{figure}
  \centering
	\includegraphics[width=0.4\columnwidth]{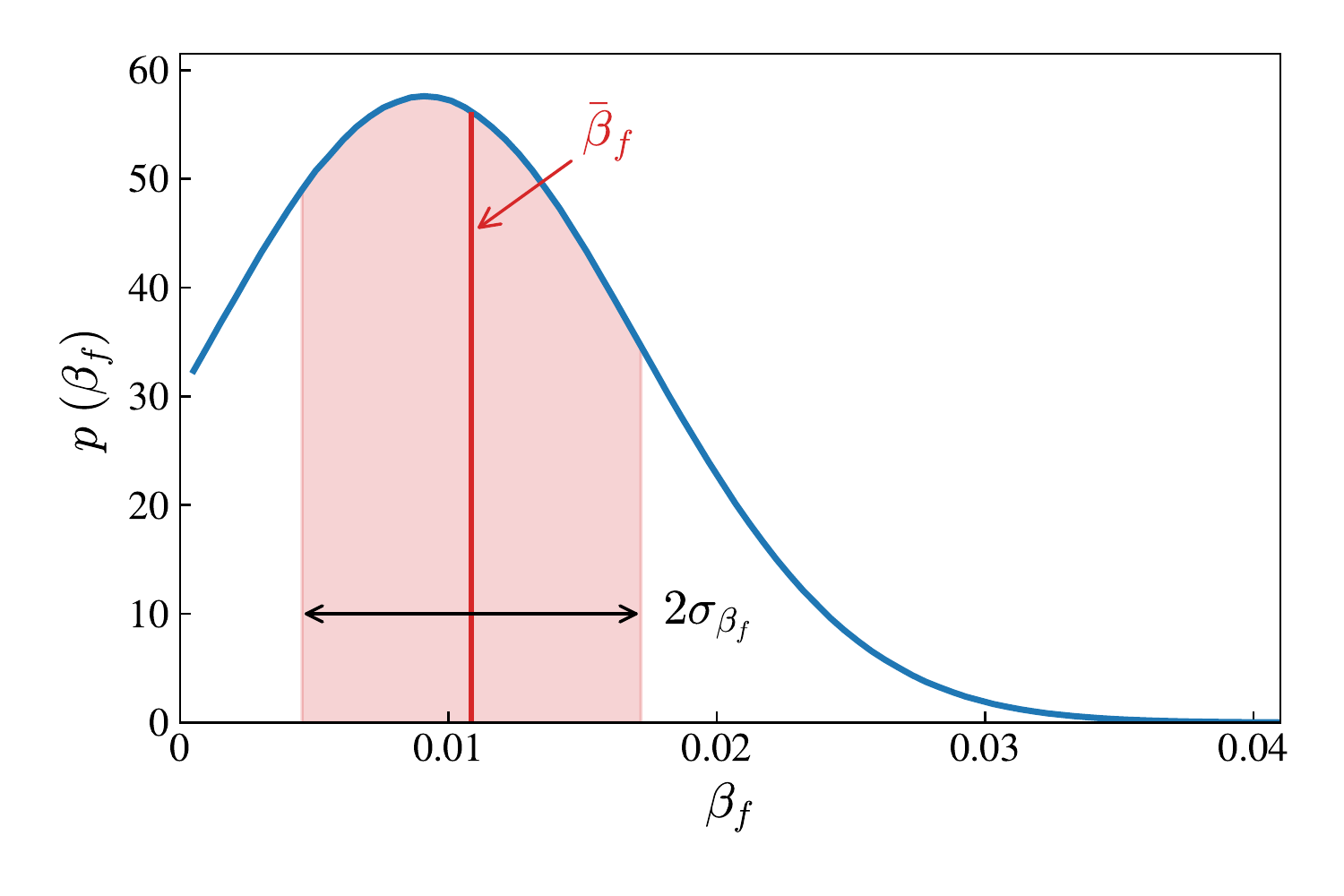}
	\caption{The blue solid line shows the probability distribution $p(\beta_f)$ that is used to model temperature-induced fluctuations of the coupling strength $\beta_f$. It is a normalized Gaussian distribution truncated at $\beta_f=0$. The fitted mean value of $\bar\beta_f=0.0108$ and the standard deviation of $\sigma_{\beta_f}=0.0065$ are shown as the red vertical line and red shaded area, respectively.}
	\label{fig:P_beta_f}
\end{figure}

The coupling strength of the atoms to the nanofiber-guided mode depends on their radial distance from the nanofiber surface. In the atomic ensemble, this distance and, accordingly, the coupling strength fluctuates due to the thermal motion of the atoms in the trapping potential. In order to model this effect, we assume a normalized truncated Gaussian distribution, $p(\beta_f)$, see Fig.~\ref{fig:P_beta_f}. We then draw a random $\beta_f$ from $p(\beta_f)$ for each atom and propagate the light field through the atomic ensemble. We repeat this procedure 100 times and average the resulting predictions for $P_f$, $\rho_{eg}$ and $\rho_{ee}$. Next, we fit the model predictions to the experimental data shown in Fig.~2(a) of the main text by performing a two-dimensional parameter sweep of $\bar\beta_f$ and $\sigma_{\beta_f}$. We find the best agreement for a standard deviation of $\sigma_{\beta_f}=0.0065$ and a mean value of $\bar\beta_f= 0.0108$, which agrees reasonably with an independent saturation measurement~\cite{vetsch2010optical} that yields $\beta_f=0.009(1)$.

\begin{figure}
  \centering
	\includegraphics[width=0.4\columnwidth]{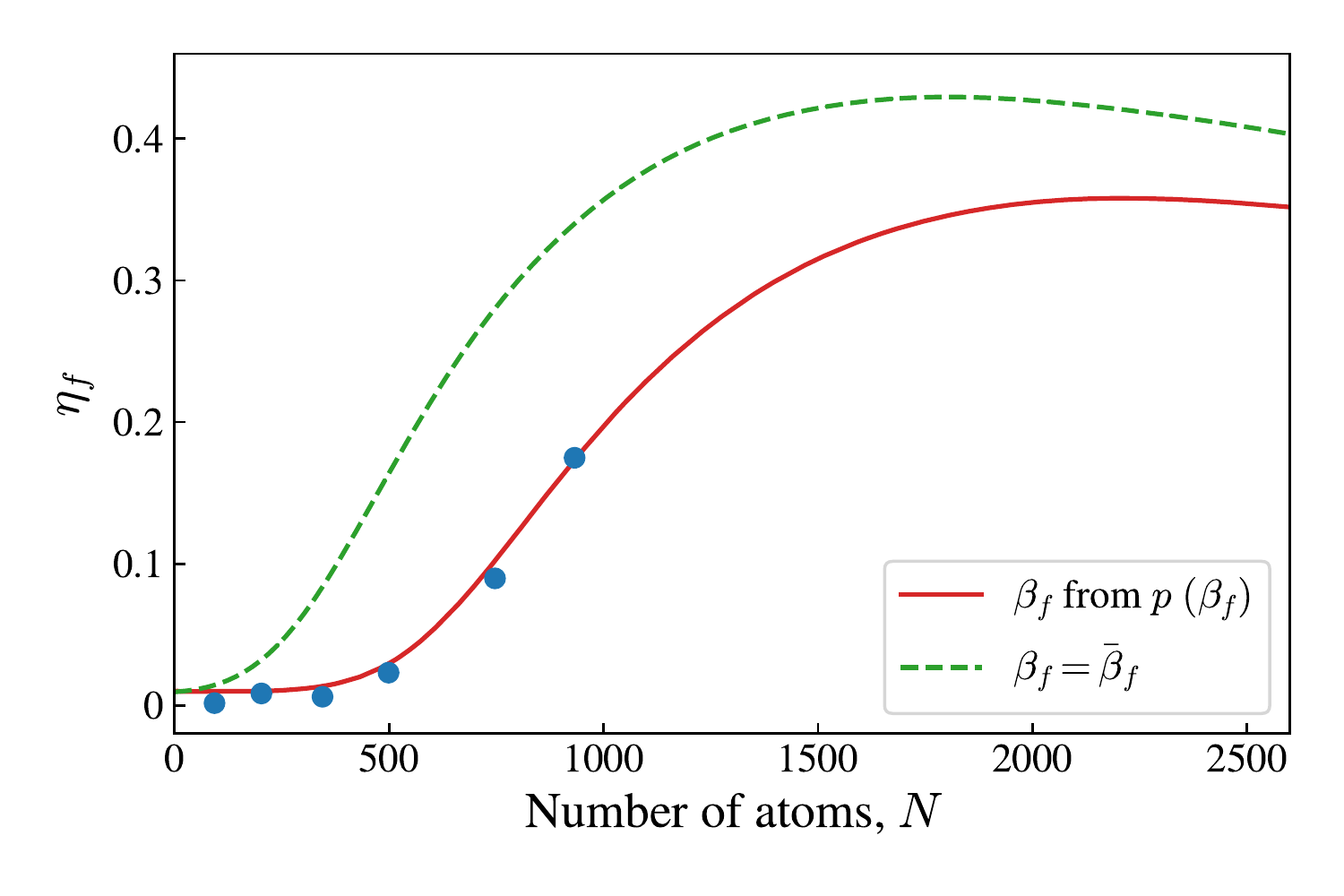}
	\caption{Fraction of stored energy emitted into the forward direction, $\eta_f$, as a function of atom number, $N$, for a pulse area of $A_1=\pi$. The blue dots show the measured data. The red solid and green dashed line show our model predictions with and without taking fluctuations of the coupling strength $\beta_f$ into account.}
	\label{fig:eta_vs_N_supplemental}
\end{figure}

Figure~\ref{fig:eta_vs_N_supplemental} shows the ratio of stored energy emitted into the forward propagating mode, $\eta_f$, as a function of atom number $N$ for a pulse area of $A_1=\pi$. The experimental data (blue dots) is in good agreement with the model predictions when assuming the fitted probability distribution $p(\beta_f)$ (red solid line). In Fig~\ref{fig:eta_vs_N_supplemental}, $\eta_f$ stays at a low value for $N<400$ and then increases superlinearly. For comparison, we show the model prediction for a fixed value of $\beta_f=\bar\beta_f$ as the green dashed line. Absent the fluctuations in coupling strength, $\eta_f(N)$ grows superlinearly even for small $N$. We note, however, that the scaling becomes eventually sublinear and $\eta_f(N)$ flattens. This happens when the number of atoms $N$ becomes comparable to the number of photons in the $\pi$-pulse (about 1300 photons).

\end{document}